\documentclass[manuscript,preprint]{aastex63}
\usepackage{amsmath}
\shorttitle{Recurrent eruptions}

\begin{document}
\title{Buildup of the Magnetic Flux Ropes in Homologous Solar Eruptions}

\author{Rui Wang}
\affiliation{State Key Laboratory of Space Weather, National Space Science Center, Chinese Academy of Sciences, Beijing 100190, China; rwang@swl.ac.cn, liuxying@swl.ac.cn}

\author{Ying D. Liu}
\affiliation{State Key Laboratory of Space Weather, National Space Science Center, Chinese Academy of Sciences, Beijing 100190, China; rwang@swl.ac.cn, liuxying@swl.ac.cn}
\affiliation{University of Chinese Academy of Sciences, Beijing 100049, China}

\author{Shangbin Yang}
\affiliation{Key Laboratory of Solar Activity, National Astronomical Observatories, Chinese Academy of Sciences, Beijing 100012, China}
\affiliation{University of Chinese Academy of Sciences, Beijing 100049, China}

\author{Huidong Hu}
\affiliation{State Key Laboratory of Space Weather, National Space Science Center, Chinese Academy of Sciences, Beijing 100190, China; rwang@swl.ac.cn, liuxying@swl.ac.cn}



\begin{abstract}
Homologous coronal mass ejections (CMEs) are an interesting phenomenon, and it is possible to investigate the formation of CMEs by comparing multi-CMEs under a homologous physical condition. AR 11283 had been present on the solar surface for several days when a bipole emerged on 2011 September 4. Its positive polarity collided with the pre-existing negative polarity belonging to a different bipole, producing recurrent solar activities along the polarity inversion line (PIL) between the colliding polarities, namely the so-called collisional PIL (cPIL). Our results show that a large amount of energy and helicity were built up in the form of magnetic flux ropes (MFRs), with recurrent release and accumulation processes. These MFRs were built up along the cPIL. A flux deficit method is adopted and shows that magnetic cancellation happens along the cPIL due to the collisional shearing scenario proposed by Chintzoglou et al. The total amount of canceled flux was $\sim$0.7$\times$10$^{21}$ Mx with an uncertainty of $\sim$13.2\% within the confidence region of the 30$^\circ$ sun-center distance. The canceled flux amounts to 24$\%$ of the total unsigned flux of the bipolar magnetic region. The results show that the magnetic fields beside the cPIL are very sheared, and the average shear angle is above 70$^\circ$ after the collision. The fast expansion of the twist kernels of the MFRs and the continuous eruptive activities are both driven by the collisional shearing process. These results are important for better understanding the buildup process of the MFRs associated with homologous solar eruptions.
\end{abstract}



\section{Introduction} \label{sec:intro}
Coronal mass ejections (CMEs) have been recognized as primary drivers of space weather. Successive CMEs have a high probability to interact with each other. A CME preconditions the pre-existing ambient solar wind for subsequent CMEs in interplanetary space \citep[e.g.,][]{2014Liuxying2,2019Liuxying}. \citet{2019Liuxying} suggest that the CMEs with the physical processes mentioned above may produce a superstorm in a ``perfect storm'' scenario, which would have a severe impact on the Earth. This hypothesis is proposed based on the contemporary observations and historical records of some extreme events. Successive CMEs that erupt from the same source region have been called homologous CMEs \citep{1984Woodgate,2002Zhangjun}, or recurrent CMEs. The research on the mechanisms of homologous CMEs eruptions is of importance for space weather forecasting, which can help to improve the prediction of extreme space weather events. In itself, the homologous CME is a kind of special CME that provides an opportunity to investigate the formation of CMEs by comparing multi-CMEs under a homologous physical condition. Therefore, it is also important for a better understanding of the physical properties of a CME itself.

Nowadays, the occurrence of most CMEs is thought to be related to a magnetic flux rope (MFR) structure above the polarity inversion line (PIL) of a CME source region \citep[e.g.,][]{1974Kuperus,1989Vanballegooijen,2010Mackay,2013Schmieder}. Regarding the formation of MFR structures, one opinion thinks that photospheric motions that are generally considered to be a result of convection and differential rotation are able to generate MFRs by tether-cutting reconnection (TCR) or magnetic cancellation \citep{1989Vanballegooijen,2001Moore}. By contrast, helical structures are thought to form below the photosphere through subsurface convection, it has been proposed that they emerge from the PIL as an MFR \citep[e.g.,][]{2008Okamoto}. \citet{2004Manchester} simulate an MFR structure emerging from the convection zone into the corona by buoyancy, generating a Lorentz force that drives shear Alfv\'en waves that transport the axial flux of the MFR into its expanding portions. They suggest that homologous eruptions would be naturally explained by the continued transport of axial flux mentioned above. Recently, \citet{2019Chintzoglou} propose that in multipolar active regions (ARs), two emerging bipoles collide at a collisional PIL, resulting in shearing and a cancellation of magnetic flux, which is also accompanied by recurring eruptions.

There have already been several studies on AR 11283 \citep[e.g.,][]{2013Fengli,2014Jiangcw,2014Ruangp,2014Liuchang,2015Romano,2015Ruangp,2015Zhangqm,2016Jiangcw,2016Dissauer,2019Sarkar}. Compared to these previous work about this AR, we are interested in how the new magnetic free energy and helicity build up from the remnant of the magnetic fields left by the previous eruption, and how the MFRs build up with the energy and helicity injected. We show that a series of eruptive events are accompanied by the cancellation process during flux emergence. This paper investigates four successive CME eruptions as a whole, which may provide a new perspective to explore the formation of CMEs.

\section{Overview of the magnetic energy and helicity of the recurrent eruptions} \label{sec:overview}
AR 11283 was located around N14E10 on 2011 September 4, having been on the solar surface for several days. We follow this AR for seven days. Four successive major eruptions happened from September 6 (two M-class flares and two X-class flares, all with CME eruption; the four eruptions are named as E1, E2, E3, and E4, hereafter). We use vector magnetic field data \citep{2014Hoeksema} from the Solar Dynamics Observatory's \citep[SDO;][]{2012Pesnell} Helioseismic and Magnetic Imager \citep[HMI;][]{2012Scherrer,2012Schou}. We adopt the Spaceweather HMI Active Region Patches \citep[SHARPs;][]{2014Bobra,2014Hoeksema} data and calculate the AR magnetic free energy by subtracting the potential field \citep[PF;][]{1978Seehafer} energy from the nonlinear force-free field \citep[NLFFF;][]{2004Wiegelmann,2012Wiegelmann} energy. The temporal resolution of the free energy is 1 hr for 120 hr. During each eruption \citep[except for E1, which will be discussed later, since no fully formed MFR structure is involved in E1 as indicated by][]{2016Jiangcw}, the free energy clearly decreases in a stepwise manner (Figure~\ref{fig:fig1}a; all exceed $\sim$10$^{32}$ erg). After each eruption, the energy is injected rapidly. During the third X1.8 flare, the amount of the free energy released reaches a maximum of $\sim$7$\times$10$^{32}$ erg, accounting for $\sim$60\% of the pre-flare free energy.

We adopt a finite volume method (FV) to calculate the relative magnetic helicity \citep{2013Yangsb,2018Yangsb}. This method entirely relies on the results of our NLFFF model and provides the helicity in a bounded volume of the corona above the AR. The overall evolving trend is similar as for the energy. There is also no obvious decrease of the helicity during E1. The helicity decreases rapidly during each of the last three eruptions due to a mass of magnetic flux ejected into the higher corona. It begins to decrease prior to the major flare of E3 \citep[the decrease is associated with a C-class flare eruption;][]{2015Ruangp}, undergoing a mild release, and even continues to decrease for a period after E3.

In order to estimate the helicity fluxes across the photosphere, we use the Poynting-like theorem derived by \citet{1984Berger} for the helicity flux through a surface. We call this method the helicity-flux integration (FI), which can be compared to the FV method to check the results \citep{2016Valori}. The vector velocity fields for the FI method are derived using the Differential Affine Velocity Estimator for Vector Magnetograms \citep[DAVE4VM;][]{2008Schuck}. These velocities are further corrected by removing the irrelevant field-aligned plasma flow as done by \citet{2014Liuyang2}. We estimate the uncertainty for each time frame of the magnetic helicity flux by conducting a Monte Carlo experiment as implemented by \citet{2012Liuyang}. First, we randomly add noise to the vector magnetic fields. Then, the noise is propagated to the velocity and helicity flux through the DAVE4VM method. The noise is subject to a Gaussian distribution, and the width of the Gaussian function is 100 G, which is roughly the noise level of the HMI vector magnetic fields \citep{2014Hoeksema}. The test is repeated 100 times for each time frame. In order to show the average temporal behavior of the magnetic helicity, the original helicity flux is smoothed using a central moving average of 2 hr time series. Therefore, we apply the same implementation to the error of the 100 experiments by the 2 hr central moving average. We use the error propagation formula below to achieve this:
\begin{equation}
(\delta \overline{H})^2 = \frac{1}{N^2}\sum_i^N(\delta H_i)^2~~~~(N = 10),
\end{equation}
where N = 10 represents the 2 hr average, i.e., the helicity taken at every 12 minute interval for each hour.

Generally, in an emerging process, the helicities that accumulate across the boundary in the corona should be equal to the increment of the volume helicity. We regard the time around 19:00 UT on September 4 as the starting time of a rapid injection of the magnetic helicity. The volume helicity increases from $\sim$--0.6$\times$10$^{42}$ Mx$^2$ to $\sim$1.3$\times$10$^{42}$ Mx$^2$ up to the beginning of E1 (Figure~\ref{fig:fig1}b). Thus, the injected helicity estimated from the FV method reaches $\sim$1.9$\times$10$^{42}$ Mx$^2$ during the rapidly emerging process. By contrast, during the period mentioned above, the accumulated helicity as measured by the FI method increases from $\sim$1$\times$10$^{42}$ to $\sim$3$\times$10$^{42}$ Mx$^2$. Therefore, the accumulated helicity of $\sim$2$\times$10$^{42}$ Mx$^2$ during this process is comparable to the helicity measured by the FV method. Figure~\ref{fig:fig1}d shows that the helicity flux from shearing motions is dominant and that its increases are connected with each eruption, which indicates that photospheric motions play an important role in the recurrent eruption process.

\section{Magnetic flux measurements and collisional shearing process} \label{sec:flux}
Figure~\ref{fig:fig2}a shows that AR 11283 was in a decay phase and that the pre-existing polarity pair P1-N1 was also decaying, but near the major polarity N1 in the south (N1$^{S}$), a pair of emerging polarities P2-N2 was tracked from the beginning of 2011 September 4 to the end of September 8. We track the proper motions of the emerging bipole via the flux-weighted centroids around the peak intensity of each polarity inside a radius of 5 Mm, which is the same as the tracking method used in \citet{2019Chintzoglou}. Referring to the animation of Figure~\ref{fig:fig2}a, the emerging bipole expands along the direction of the yellow arrows shown in Figure~\ref{fig:fig2}a. The dynamic evolution of the emerging bipole also can be expressed in a 3D representation by means of an image stacking method \citep{2013Chintzoglou}. The 3D space-time representation of the magnetic flux tube structures in Figure~\ref{fig:fig2}b indicates the connectivity of the pairs of magnetic polarities.

\citet{2019Chintzoglou} proposed a ``Flux Deficit'' method to measure the canceled magnetic flux around the PIL in the flux emergence process. This method calculates the deficit of the imbalanced flux of a pair of conjugated polarities to estimate the canceled flux around a collisional PIL (cPIL) between two unconjugated polarities. We calculate the amount of magnetic flux as $\Phi = \int B_r dA$, where B$_r$ is the radial component of the photospheric magnetic field vector from the SHARP data set. For the HMI vector data set, the noise level for B$_r$ is estimated to be that of $|B|~\sim$ 100 G \citep[1$\sigma$ level;][]{2014Hoeksema}. They indicate that calculating the flux from B$_r$ can lead to a spurious increase of the flux toward the limb, especially for the locations $\Theta$ $\gtrsim$ 30$^{\circ}$ from the disk center. Therefore, we try to use the B$_r$ data within the sun-center distance of 30$^\circ$ to mitigate this effect.

We follow the method of \citet{2019Chintzoglou} to extract the location of the cPIL by performing the gradient operation on the magnetograms described in \citet{2007Schrijver}. We use a threshold of 100 G for B$_r$ in calculating the gradient for comparison with the results in \citet{2019Chintzoglou}. Furthermore, we dilate the bitmaps of positive and negative polarities and get an overlap of $\leq$ 5 pixels width. Next, we multiply the bitmap of the overlap by B$_r$, and use contours to get the cPIL. The animation of Figure~\ref{fig:fig2}a shows that there are some breaks of the cPILs. It indicates that the polarities are not compact, which is due to the dynamic evolution of the emerging fluxes. As a result, the polarities around the PILs are not always associated with strong fields and high gradients. By contrast, we contour B$_r$ where $|B|~\geq~100$ G and get the longest ``zero'' level contour, the self-PIL (sPIL) between the conjugated polarity pair P2-N2. Due to the discrete new fluxes emerging between P2 and N2, the sPIL has a winding shape and changes its location with the emergence.

To implement the flux deficit method, we have to calculate each of the conjugated polarities individually. A mask is needed first to separate the like-signed polarities from different bipoles. There are some rules to define the mask (Chintzoglou 2021, private communication). (1) We must properly isolate the individual polarities from other like-signed polarities, and (2) we must contain as much surface area as necessary to account for the dispersion of flux from the polarity concentrations to the quiet Sun areas. We assume that the flux decay rate is similar for each polarity in our study, and therefore the rate of escape of flux from each mask is assumed to be the same. (3) We allow masks for conjugated polarities (opposite polarities can be distinguished by setting thresholds) to overlap, but no overlap of masks for like-signed polarities from different bipoles is allowed. (4) When measuring the flux of a conjugate bipole, i.e., P2-N2, the masks for positive and negative polarities should have equal pixel areas, to mitigate the errors coming from summing different amounts of quiet sun areas around the polarities to ensure a flux balance. For our case, P2 and N2 do not move too fast, and as a result it is easy to define a large area in which the negative and positive fragments can be counted in the P2 and N2 fluxes within the selected areas, respectively (see Figure~\ref{fig:fig2}a and the related animation).

According to the masks above, we give the flux plots in Figure~\ref{fig:fig3}. It shows that P2-N2 started to emerge around 09:00 UT September 4. In fact, the emerging bipole flux is not balanced (there is an excess negative flux bias of about 6 $\times$ 10$^{20}$ Mx). One part of the excess negative flux comes from the pre-existing flux belonging to N1$^S$, and the other part belongs to a bipole emerging along the north-south direction (shown by the purple ellipse in Figure~\ref{fig:fig4}a). Since P2-N2 entered a fast emergence phase and the flux monotonically increased after 09:00 UT September 4, we remove the bias from the measurements of the N2 flux, for consistency we correct both N2 and P2 fluxes by subtracting their pre-existing flux bias from the measurement at 09:00 UT. The correction for the bias uses the same methodology as in \citet{2019Chintzoglou}. Then, their fluxes are almost balanced and keep increasing until around 03:00 UT September 5 (see Figure~\ref{fig:fig3}). We call this phase the first emergence episode, during which a rapid emergence of new flux is ongoing (see Figure~\ref{fig:fig4}b). In this episode, the flux of the positive polarity exceeds the negative one for awhile, which is likely due to the emergence of the north-south direction bipole, but soon they decay or are canceled. Then, P2 and N2 return to balance.

We try to determine the onset of collision via the cPIL's length, as done in \citet{2019Chintzoglou}. However, we note that the configuration of the cPIL is a little different from those in \citet{2019Chintzoglou}. For their cases, the cPIL's length grows from zero when the unconjugated polarities begin to touch each other. But for this case, due to the pre-existing north-south direction bipole (Figures~\ref{fig:fig4}a and~\ref{fig:fig4}b), the initial length of the cPIL is hardly determined. More specifically, it is hard to distinguish the cPIL between P2 and N1$^S$ from the pre-existing sPIL of the north-south direction bipole at the beginning of the second emergence episode. Nevertheless, we find that it is not impossible to determine the cPIL's length, and we can visually determine the length from the magnetogram observations. Figure~\ref{fig:fig5} shows that after the first rapid emergence episode, P2-N2 separate asymmetrically. From 20:00 UT September 4 to 08:00 UT September 5, N2 moves southwestward but P2 almost stays in almost the same place (refer to the gray vertical dotted guide lines in Figure~\ref{fig:fig5}a and~\ref{fig:fig5}b). After that, P2 starts to move eastward, and the southern part of P2 begins to push N1$^S$ and forms a relatively continuous PIL (see Figure~\ref{fig:fig5}c). Therefore, we regard 08:00 UT September 5 as the onset of the converging motion of the two colliding polarities. In order to calculate the cPIL's length, we try to extract the PIL close to the converging motion as possible (as shown in the pink dashed squares in Figure~\ref{fig:fig5}b and~\ref{fig:fig5}c, in which we think the real collision process happened). For comparison, we define the onset of collision when the cPIL's continuous length definitely exceeds 40 Mm, which is the same as in the manipulation of \citet{2019Chintzoglou}. Figure~\ref{fig:fig6} shows that the onset of collision is around 18:00 UT September 5, i.e., during the second emergence episode. The large fluctuation of the length in Figure~\ref{fig:fig6} is due to the dynamic discrete emerging fluxes (see Figure~\ref{fig:fig4} and~\ref{fig:fig5}), which sometimes prevent us from determining a continuous and smooth cPIL. Therefore, we remove the outliers and smooth the curve of length.

After the onset of collision, the first major eruption occurrs at on 01:35 UT September 6. Meanwhile, the flux deficit between N2 and P2 starts to monotonically increase (see Figure~\ref{fig:fig3}). The cPIL becomes more continuous and the length of the cPIL stays above 40 Mm. Figure~\ref{fig:fig4}d shows that the southern arm of P2 (blue short arrow) push N1$^S$ eastward. A negative parasitic polarity N1$^P$ emerges (as shown by the enlarged view) and begins to move westward. The positive counterpart P1$^P$ of N1$^P$ is dispersed around N1$^P$ and cancels with N1$^N$, which results in the reduction of N1$^N$ (compare the N1$^N$ in Figures~\ref{fig:fig4}d with~\ref{fig:fig4}f). However, the reduction of N1$^N$ is not completely caused by the cancellation with P1$^P$. The eastward-moving P2 also cancels with N1$^N$. The cyan dashed ellipse in Figure~\ref{fig:fig4}e shows that N1$^P$ gradually merges with N1$^S$. The parasitic bipole is located around the unconjugated polarities P2 and N1 (between N1$^N$ and N1$^S$). It is difficult to isolate P1$^P$ from P2 in a regularly shaped mask. Therefore, we group P1$^P$ with P2. We determine that P1$^P$ accounts for only $\sim$1\% of the flux of P2. The flux of P1$^P$ dynamically changes during the emergence of the parasitic bipole, and meanwhile it continuously cancels with N1$^N$, which keeps the contribution of P1$^P$ to the P2 flux low. The merged flux inherits the movement of N1$^P$ and moves westward. The red arrow in Figure~\ref{fig:fig4}f shows that the westward merged flux sheared with the eastward-moving P2. A strong shearing motion happens along the cPIL. Figure~\ref{fig:fig3} shows that P2-N2 moves into the third emergence episode around 16:00 UT September 6. We can observe that the new flux emerges around the sPIL in Figure~\ref{fig:fig4}e; (see the animation of Figure~\ref{fig:fig2}a). At the phase prior to E3, the center of the region of interest moves out of the 30$^\circ$ sun-center distance. Figure~\ref{fig:fig4}g shows that N1$^{S}$ is approaching the mask of N2, which is not allowed by our rule that prhibits the overlap of masks for like-signed polarities from different bipoles. Flux distributions become complicated in the region due to the significant flux decay from the polarity concentrations. N2 and the westward-moving N1$^S$ become difficult to separate. Separation would introduce large uncertainties to our measurements, especially for the location $\Theta~\gtrsim~30^\circ$. In spite of this, Figure~\ref{fig:fig4}g and~\ref{fig:fig4}h show that the majority part of the westward-moving polarities (shown by the red arrows) never get into the N2 mask. Although we do not consider this part of the flux deficit after September 8, the deficit probably continues to increase with the ongoing flux emergence.

The flux measurements show that the maximum deficit within a confidence region of the sun-center angle ($\leq$ 30$^\circ$) is $\sim$0.7 $\times$ 10$^{21}$ Mx (see the green curve in Figure~\ref{fig:fig3}), which is comparable to that of AR 11158 in \citet{2019Chintzoglou}. The uncertainty comes from three sources, which are estimated following the method as in \citet{2019Chintzoglou}. The first source is the choise of threshold $|B|$. When a threhold $|B|$ = 100 G (1$\sigma$ noise level) is set, it reduces the maximum deficit by $\sim$1.8\% within the 30$^\circ$; if set $|B|$ = 200 G (2$\sigma$ noise level) is used, it reduces the maximum deficit by $\sim$51.9\% within the 30$^\circ$. This shows that the deficit decreases quickly when the noise level is changed from 100 to 200 G. Our explanation for this is that the leading positive polarity P2 is more concentrated, and the following negative polarity is more dispersed, which is visualized by the 3D space-time representation of the magnetic flux tube structures in Figure~\ref{fig:fig2}b. When the threshold is increased to some value between the dispersed negative flux and the stronger positive flux, more dispersed negative fluxes are removed from the total N2 than from the positive fluxes, which reduces the flux deficit largely. Thus, the source of uncertainty, u$_{noise}$, comes from trying to reduce the instrument noise in the SHARP data for the flux deficit calculation.

A second source of uncertainty is from flux imbalance (u$_{imbalance}$) caused by the flux imbalance of P2 and N2 (i.e., P2 $\geq$ N2) before the onset of collision (N2 $\geq$ P2 is reasonable). The imbalance reduces the maximum deficit by $\sim$13.0\%. The imbalance is probably caused by the new emergence of the positive flux from the north-south bipole mentioned above.

The third source of uncertainty comes from difficulties in grouping polarities. As we have already determined the P2 and N2 are balanced at 09:00 UT September 4, the south-north bipole (see Figure~\ref{fig:fig4}a) is considered to be part of the unconjugated polarities P2 and N1$^S$. Even if they continued to emerge after the initiation, we would not consider them to be a conjugated bipole. The uncertainty from the flux imbalance associated with new flux emergence has been considered in the second source; thus we will not count this in the third source. Regarding the positive parasitic polarity P1$^P$, although we group it with P2, it gradually cancels with N1 and does not exist after N1$^P$ merges with N1$^S$ on September 6 (see Figure~\ref{fig:fig4}d and~\ref{fig:fig4}e). Thus, the influence of the parasitic polarities on the ``grouping'' uncertainty would not be considered. By contrast, Figure~\ref{fig:fig4}g shows the east edge of the N2 mask around the sPIL is very close to N1$^S$, which is not allowed by rule (3). We determine this part of the ``grouping'' uncertainty by adding or removing magnetic elements from the mask. The uncertainty is small (only $\sim$1.5\%) at the maximum deficit within the confidence region, but with the westward-moving polarity approaching the N2 mask and gradually mixing with N2 (see Figure~\ref{fig:fig4}g and~\ref{fig:fig4}h), the uncertainty increases rapidly, reaching $\sim$68.2\% at 00:00 UT September 8. As the AR has already moved out of the confidence region (N2 should be located further west) and also has such a large uncertainty, this part of the uncertainty after E3 will not be taken into consideration. Thus, the third source of uncertainty (u$_{grouping}$) is $\sim$1.5\% within the confidence region.

When the AR gets complicated on September 8 (see Figure~\ref{fig:fig4}g and~\ref{fig:fig4}h), it has moved too far beyond the 30$^\circ$ distance from the disk center. This is similar to the situation of AR 11158 in \citet{2019Chintzoglou}. N2 and the westward-moving N1$^S$ became hard to separate; such a procedure would lead to large uncertainties in our measurement. Therefore, the total uncertainty for AR 11283 is

\begin{equation}
\begin{aligned}
u_\Delta^{AR11283} &= \sqrt{u_{noise}^2+u_{imbalance}^2+u_{grouping}^2} \\
&= \sqrt{(1.8\% \sim 51.9\%)^2+(13.0\%)^2+(1.5\%)^2} = 13.2\% \sim 53.5\%~(1\sigma \sim 2\sigma).
\end{aligned}
\end{equation}

\section{Buildup of the MFRs} \label{sec:MFR}

As shown in Figure~\ref{fig:fig1}a and~\ref{fig:fig1}b, E1 is not associated with an MFR eruption above the major PIL, which is also shown by \citet{2016Jiangcw}. The first row in Figure~\ref{fig:fig7} shows the twist map based on our extrapolation results on a cross section \citep{2016Liurui} shown in Figure~\ref{fig:fig2}a. The left and right maps (before and after E1) show that there is no obvious change in the kernel fields above the PIL. By contrast, the other three events (the left column) are all associated with a strong twist kernel on the cross section. We consider the kernel with strong twists as the core of the MFR structures. It shows that the cores disappears after each of the last three eruptions (the right column) and the twists of the MFR's envelop fields also decrease.

We trace a rectangular region and make sure that its long side is always parallel to the PIL (see the green rectangle in Figure~\ref{fig:fig2}a). The upper panel of Figure~\ref{fig:fig8} presents the time evolution of the tangential component of the photospheric magnetic fields (B$_t$) within the region. The increase of B$_t$ during each eruption is consistent with the conjecture proposed by \citet{2008Hudson}, which suggests that the magnetic loops should undergo the implosion phenomenon due to the energy release processes. The gradual decrease of B$_t$ after each eruption probably manifests such that the magnetic loops are stretched upward when the coronal magnetic pressure increases with newly supplemented energy (see Figure~\ref{fig:fig1}a). We also extract the component of B$_t$ parallel to the PIL (B$_{tpara}$). The black and red curves suggest that most of the component of B$_t$ is along the direction parallel to the PIL, which indicates that the fields after the onset of collision (18:00 UT September 5) become more sheared and part of nonpotentiality remains along the PIL after the eruptions. B$_{tpara}$ approaching to B$_t$ indicates the collision process of P2 and N1$^S$. Actually, the field lines along the PIL became very sheared from 16:00 UT September 6 (i.e., the curves of B$_t$ and B$_{tpara}$ become very closed), when a spot rotation occurrs (see the later text), they remain very sheared after that.

In order to figure out how sheared the fields are, we calculate the magnetic shear angle as done by \citet{1992Wanghm}, \citet{2010Gosain}, and \citet{2012Petrie}, among other authors. The shear angle $\Delta\phi$ is between the observed and potential field \citep{1978Seehafer} azimuths and is defined by
\begin{equation}
\Delta\phi = cos^{-1}\frac{\mathbf{B}_t^o \cdot \mathbf{B}_t^p}{|B_t^o||B_t^o|},
\end{equation}
where $\mathbf{B}_t^o$ and $\mathbf{B}_t^p$ are the observed and potential horizontal field vectors. The magnetic shear is averaged within the green rectangular region shown in the animation of Figure~\ref{fig:fig2}a. The green curve in the lower panel of Figure~\ref{fig:fig8} shows that the shear angle increases from the onset of collision and stops increasing until September 8, when the average shear angle reaches $\sim$78$^\circ$. The fast growth phase of the magnetic shear is prior to September 7. Figure~\ref{fig:fig4}d shows that P2 is moving eastward, but the fields of P2 are not completely antiparallel to its direction of motion around the beginning of September 6. P2 gradually intrudes into the surrounding negative fluxes consisting of N1$^N$ and N1$^S$ on September 6 (see the animation of Figure~\ref{fig:fig2}a). After the later time of September 6, the fields of both P2 and N1$^S$ become almost completely antiparallel or parallel to their moving directions, and the growth of the shear angle slows down. P2 and N1$^S$ gradually turn into a full shearing motion on September 7.

An MFR is the key structure related to the major eruptions. Its evolution is crucially important for understanding the mechanisms of the eruption. The top four panels in Figure~\ref{fig:fig9} show the decay index equal to 1.5 with a height contour overlaid on photospheric magnetograms for the four major eruptions. The decay index, $n = -\partial {ln B_h}/\partial {ln z}$, characterizes the rate of decrease of the horizontal component of the potential field B$_h$ with height $z$. For bipolar configurations and for n$\geq$1.5, an MFR is torus-unstable and thus cannot be constrained by the overlying field. The areas within the color contours are called ``super-critical'' decay indices by \citet{2015Chintzoglou}. The continuous stacking of the ``super-critical'' decay index areas gives birth to a ``super-critical'' decay index ``tunnel,'' which is favorable for the escape of a CME. The background magnetic structures are changing with the photospheric motions, resulting in the change of the strapping force above the MFR. On September 6, the tunnel is located along the PIL of P2-N1$^N$, and gradually it moves to the PIL of P2-N1$^S$. The bottom panel in Figure~\ref{fig:fig9} also presents the time evolution of the critical height for torus instability with an average decay index equal to 1.5, which is calculated within the yellow rectangular region shown in Figure~\ref{fig:fig2}a. The critical height changes between 13 and 25 Mm and reaches its minimum around E3. The value of the critical height is relatively low compared with those of other eruptive events \citep[e.g.,][]{2015SunXD}.

Figure~\ref{fig:fig10}a shows the time evolution of the height of each MFR. Here, we assume that the axis of an MFR is located at the twist-weighted centroid of a strong twist kernel (twist number $>$ 1; see Figure~\ref{fig:fig7}) on the twist map, and we regard the height of the twist-weighted centroid as the height of each MFR \citep[e.g.,][]{2014Jiangcw,2020Hewen}. The height presents an increasing trend before each of the last three eruptions (see the red dots in Figure~\ref{fig:fig10}a).

Interestingly, the rise of each MFR exhibits a simultaneous change with the photospheric motions. When the polarities P2 and N1$^S$ collide with each other, caused by the separation of P2-N2, the sunspot rotation of P2 starts around 14:00 UT September 6 (Figure~\ref{fig:fig10}b). We present an $r - \theta$ stackplot in Figure~\ref{fig:fig10}c. The stackplot is produced by stacking an arc slit of 150$^\circ$ with a radius of $\sim$3$^{\prime\prime}$.5. The stackplot presents the variation of the rotation speed by tracking the magnetic arm in the north of P2. The angle $\theta$ increases in the clockwise direction from $-$50$^\circ$ to 100$^\circ$ with $\theta = 0^{\circ}$ along the upward direction (the north). \citet{2014Ruangp} has gotten similar results on this spot rotation using continuum images. The highest value of the magnetic arm (red line) describes an acceleration phase from 14:00 UT to 18:00 UT September 6, which just corresponds to the rise of the MFR (see Figure~\ref{fig:fig10}a). After 19:00 UT, the rotation speed shows a further acceleration that corresponds to a faster rise of the MFR. There are obvious changes in both the magnetic energy and helicity after 15:00 UT, as well as in the flux deficit by the cancellation (see Figure~\ref{fig:fig3}). The shear term in Figure~\ref{fig:fig1}d also shows a simultaneous rapid increase with the acceleration of the sunspot rotation.

The height of the MFR on September 7 presents a relatively uniform growth. Figure~\ref{fig:fig10}d shows the shearing motions between P2 and N1$^S$. The length of the separation in Figure~\ref{fig:fig10}e (thick blue line) shows that the shearing motions in the east-west direction continue from E2 to E3. P2 and N1$^S$ start to converge to each other around 12:00 UT (red line) and then stop until the later stage of the eruption. The flux deficit P2 and N1 have an obvious additional increase during this period. It implies that the converging motion enhances the collisional shearing process. We notice that the shear term of the helicity flux in Figure~\ref{fig:fig1}d shows a rapid change from 08:00 to 12:00 UT. Correspondingly, Figure~\ref{fig:fig10}a shows that a strong twist kernel appears during this period, which enables us to trace the height of the MFR. Then, the MFR enters a fast growth period during the converging time (red line). Around 20:00 UT, the height of the MFR descends. This is probably because that part of the rope has already erupted into the higher corona, which may be associated with the extreme ultraviolet (EUV) observations of \citet{2015Ruangp}. The helicity determined by the FV method supports our idea; it obviously decreases after 18:00 UT. However, the energy seems inconsistent with the helicity. Figure~\ref{fig:fig3} shows that the fluxes of P2 and N2 both emerge from 14:00 UT on September 7, which probably compensates for the released energy. After E3, P2-N1$^S$ enters a faster shearing phase (green line). Correspondingly, the MFR evolves faster than the previous two. Compared with the previous two eruptions, we can trace the strong twist kernel in an earlier time just after E3. The flux deficit also has a fast growth during this period. However, due to the difficulties in grouping polarities, the uncertainty increases greatly to $\sim$69.6\% at the beginning of September 8, during which time the AR has just moved out of the confidence region of the 30$^\circ$ sun-center distance. Therefore, the reliability of the results is reduced. Nevertheless, the consistent behaviors described above indicate that there is a strong correlation between the photospheric motions and the growth of the MFR.

\section{Discussion and summary} \label{sec:disc}
The homologous events in AR 11283 present an integrated physical process showing how a new flux system emerges into a mature AR and interacts with the pre-existing ambient fields, causing a series of successive eruptions. The so-called ``collisional shearing'' process proposed by \citet{2019Chintzoglou} happens between the emerging flux and the pre-existing ambient fields. It is more difficult to determine the accurate onset time of collision than for the cases in \citet{2019Chintzoglou}. Due to the pre-existing bipole along the north-south direction (see Section 3), it is hard to distinguish the cPIL between the unconjugated polarities P2 and N1$^S$ from the pre-existing sPIL of the north-south direction bipole. We try to follow the method of \citet{2019Chintzoglou} to determine the onset of collision via the cPIL's length, but due to the reason mentioned above, the first thing we need to do is to determine the cPIL's length as accurately as possible. It is found that the second emergence episode is very important for the collision. We can visually distinguish the cPIL from the pre-existing sPIL via the magnetogram observations, and then use the same criterion as in \citet{2019Chintzoglou} to determine the onset time of the collision. P2 starts to move eastward around 08:00 UT on September 5 and begins to push N1$^S$ eastward. The strong flux emergence phase of P2-N2 appears from 16:00 UT September 5 to the onset of September 6, during which time the continuous cPIL reaches a continuous length of $\geq$ 40 Mm around 18:00 UT September 5 (see Figure~\ref{fig:fig5}b), and P2 deforms N1$^S$ such that it becomes an elongated shape along the cPIL. The results confirm that the onset of collision is during the second emergence episode.

\begin{figure}
\plotone{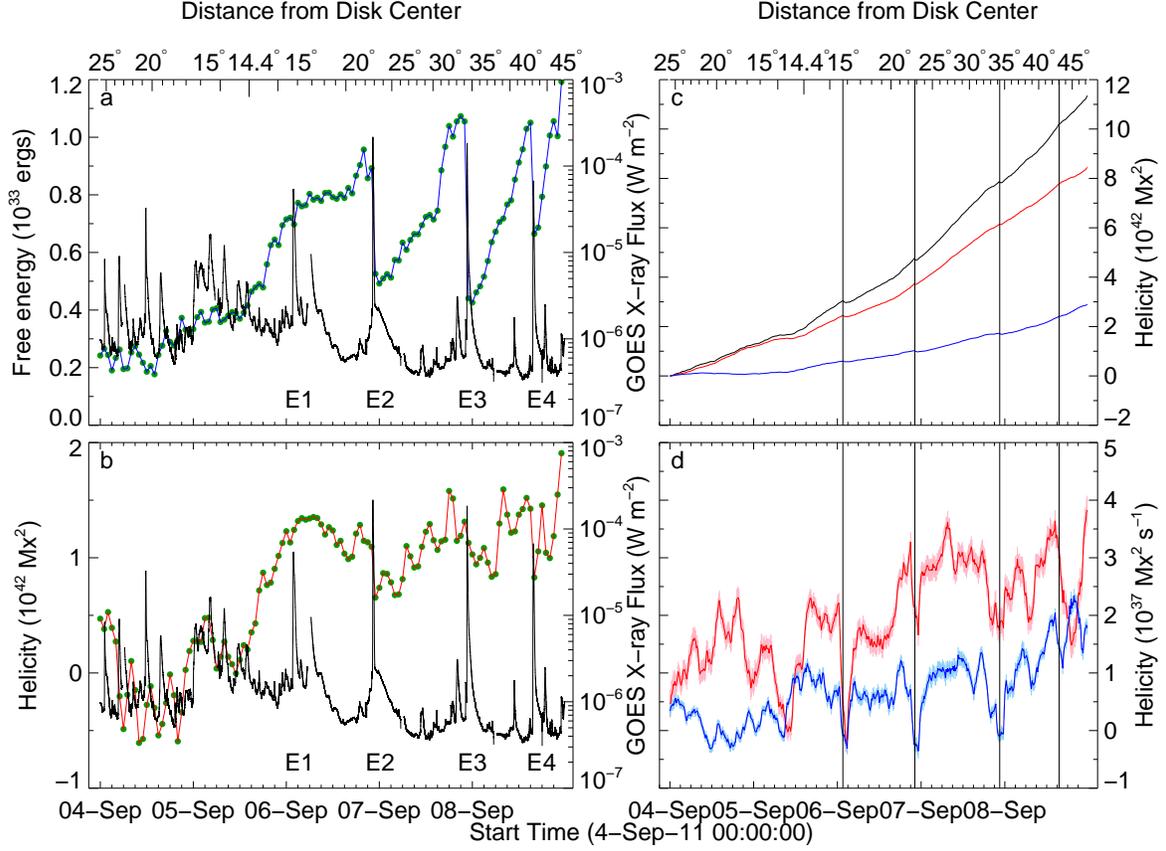}
\caption{Evolution of magnetic free energy and helicities of AR 11283 over five days. (a) Magnetic energy (blue) derived from the NLFFF and PF extrapolation with an 1 hr cadence (green dotted line). (b) Magnetic helicity calculated by the finite volume method. GOES soft X-ray flux (1--8 \AA~channel, black) is overplotted in (a) and (b). (c) Accumulated helicities from the shear term, the emergence term, and the total are colored in red, blue, and black, respectively. (d) Helicity fluxes across the photosphere from shear and emergence terms are colored in red and blue, respectively. The results are smoothed using central moving averages of 2 hr time series. The 1$\sigma$ error is presented by the light colored error bars. The sun-center angle of the emerging region is presented on the top $x$-axes. The vertical lines represent the onset times of the successive eruptions.\label{fig:fig1}}
\end{figure}

\begin{figure}
\plotone{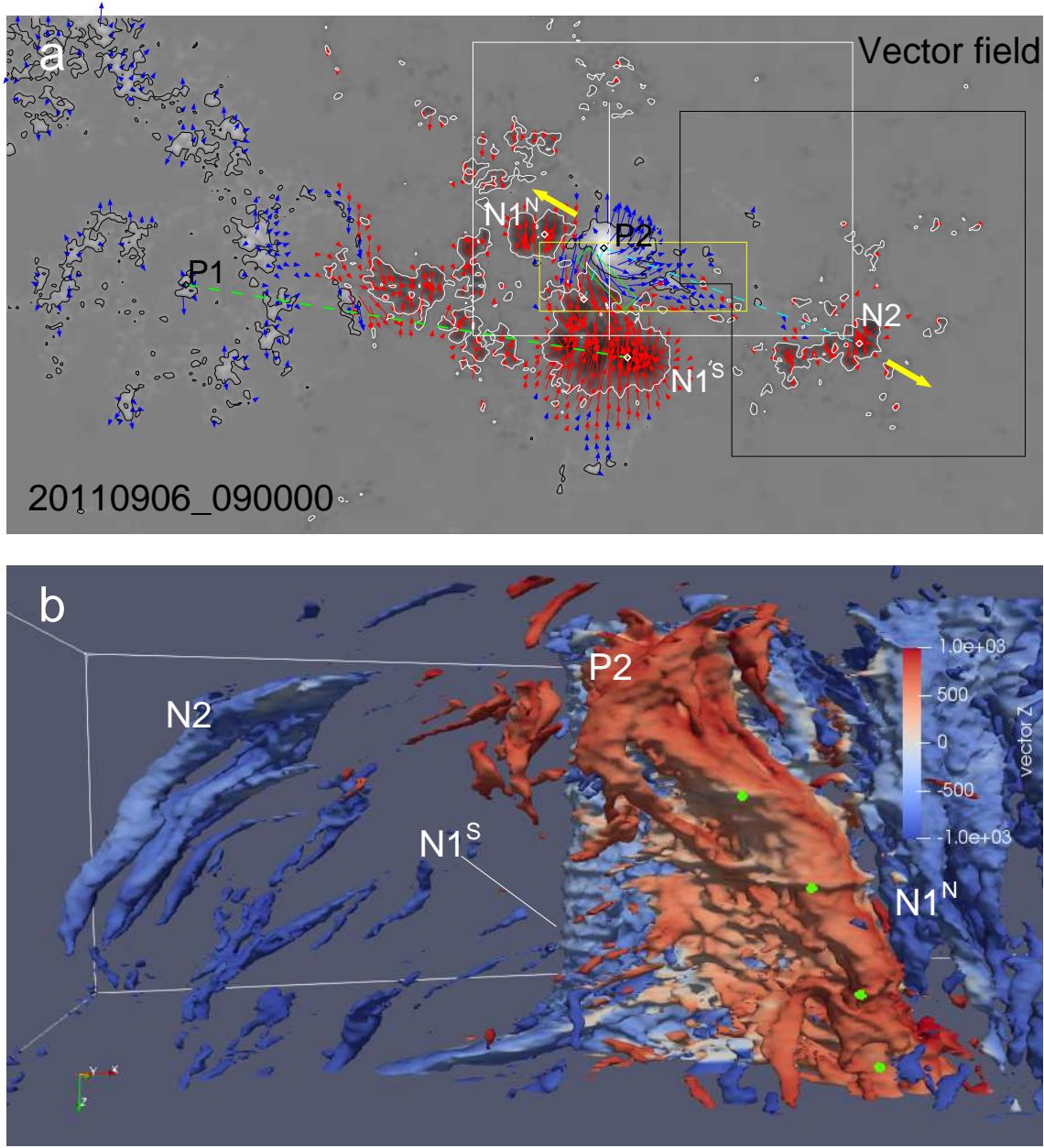}
\caption{(a) SHARP vector magnetogram for the AR 11283. The vertical field (B$_z$) is plotted in the background with isocontours at 350 G ($-$350 G) in black (white); blue (red) arrows indicate the transverse field (B$_t$; $|$B$_t| >$ 200 G) with a positive (negative) vertical component. The yellow and green rectangles indicate the regions for calculating the average decay index and the shear angle along the collisional PIL, respectively. The yellow arrows present the moving directions of P2-N2. The white rectangle and the black polygon show the masks for calculating the magnetic fluxes of P2 and N2, respectively. An animation of this figure is available. The animation starts on 2011 September 4 at 00:00:00 UT and ends on 2011 September 8 at 23:48:00 UT. The real-time duration of the animation is 40 s. (b) 3D spacetime representation \citep{2013Chintzoglou} of the emerging bipoles using isosurfaces of $|\textbf{B}|$ = 1000 G as viewed from the north. The green dots show the times of the four eruptions.}\label{fig:fig2}
\end{figure}

\begin{figure}
\plotone{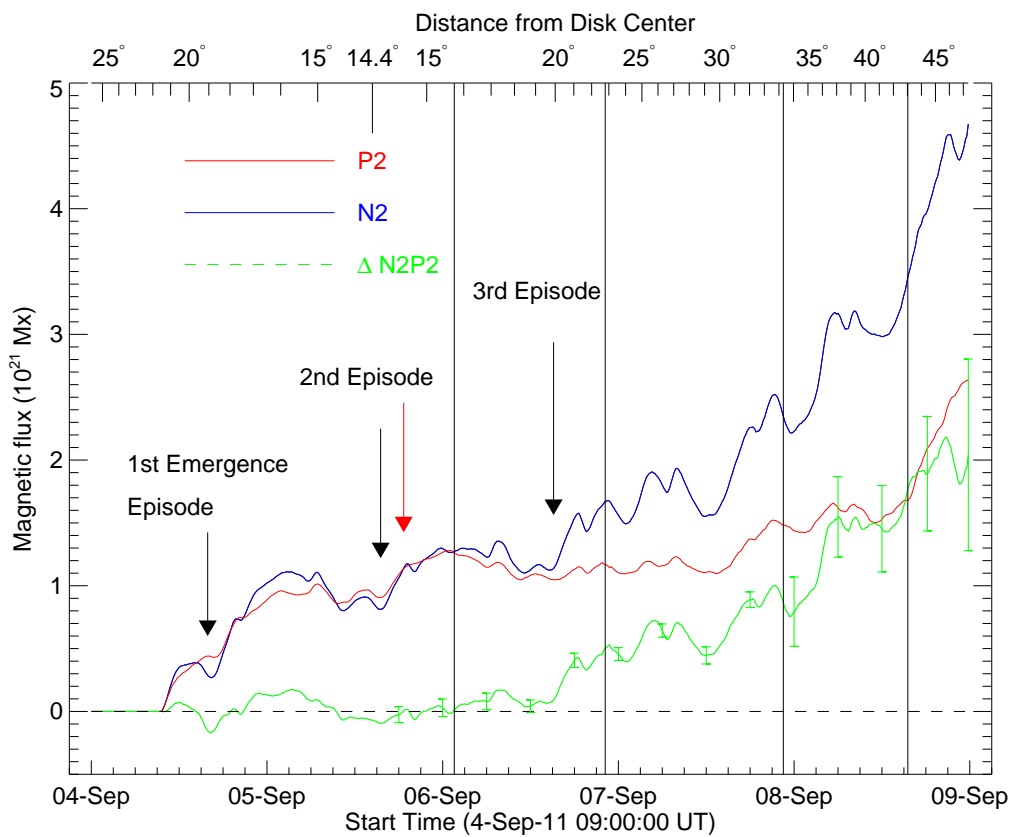}
\caption{Time evolution of the magnetic flux for the conjugated polarities P2 (red) and N2 (blue). The green curve represents the flux deficit of N2-P2. The sun-center angle of the emerging region is presented on the top $x$-axis. The black arrows indicate the onset time of each flux emergence episode. The red arrow shows the onset of collision. Error bars for the deficit are plotted every 6 hr from the onset of collision. The vertical lines mark the onset times of the four eruptions.} \label{fig:fig3}
\end{figure}

\begin{figure}
\plotone{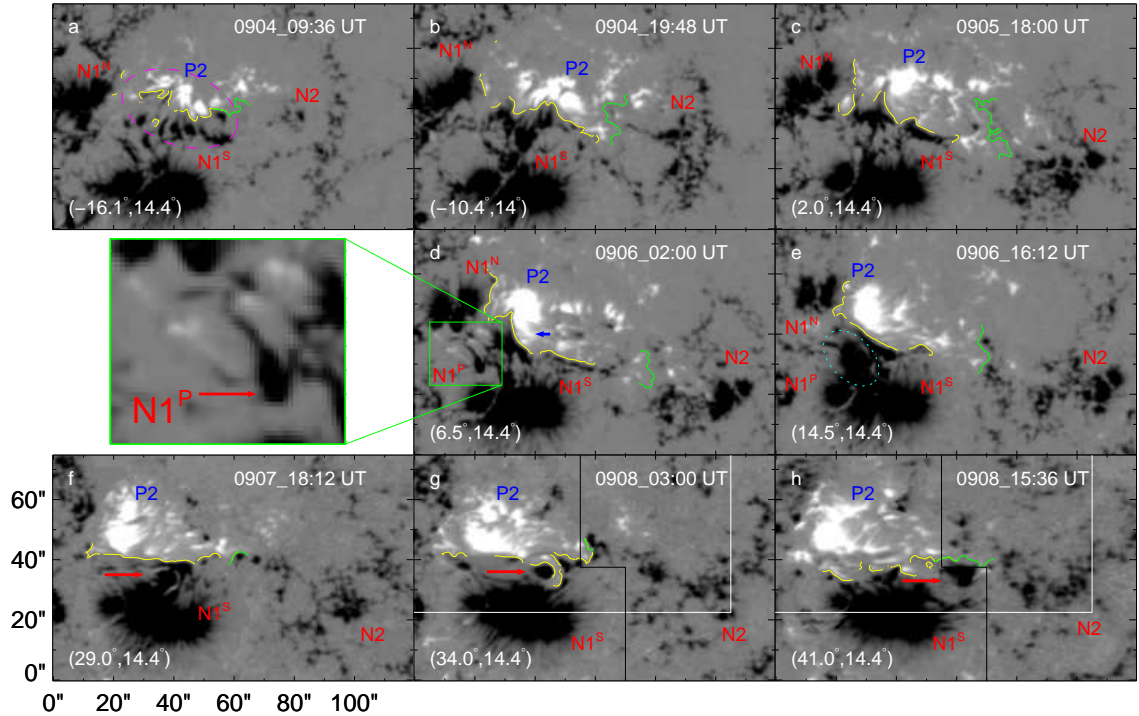}
\caption{Representative snapshot magnetograms showing the evolution of AR 11283 as observed by HMI on the Solar Dynamics Observatory. The separating conjugated polarities are annotated as P2-N2. The self-PIL of P2-N2 is represented by the green curve. Only the negative polarities N1 (including N1$^S$ and N1$^N$) of P1-N1 surrounding P2 are presented. The collisional PIL between P2 and N1 is represented by the yellow curve. The north-south direction bipole is shown by a purple dashed ellipse in (a). The green box in (d) shows the parasitic polarities with an enlarged view on the left. The cyan dashed ellipses in (e) show the negative parasitic polarity (N1$^P$) which merges with N1$^S$ and moves westward. The red arrows in (f), (g), and (h) indicate the merged negative polarity and its direction of motion. The black and white lines in (g) and (h) are the partial mask regions shown in Figure 2a. The grayscale magnetograms are saturated at $\pm$ 1000 G. The coordinate of the disk center is also given in each panel.} \label{fig:fig4}
\end{figure}

\begin{figure}
\plotone{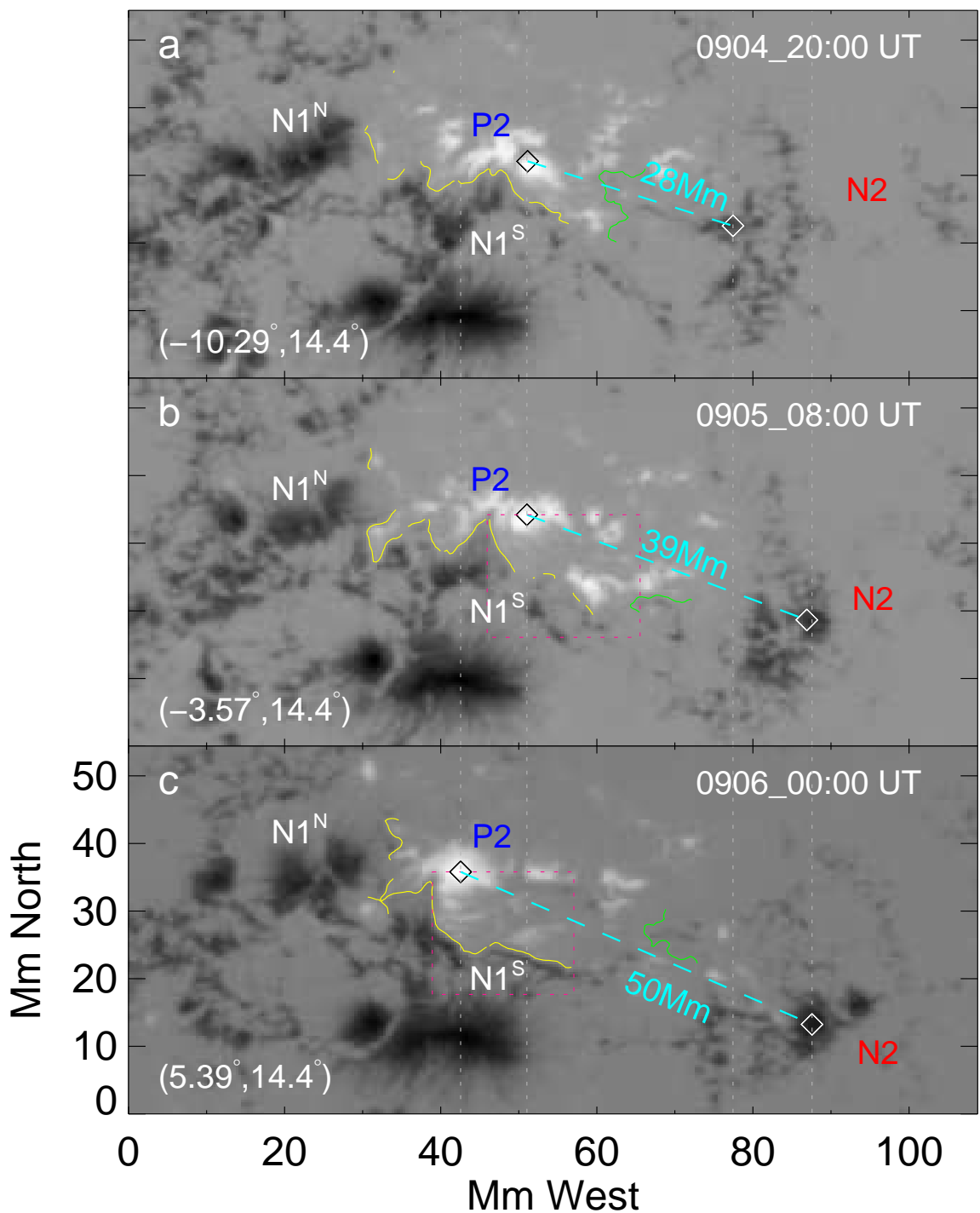}
\caption{Asymmetrical separation of the conjugated polarity pair P2-N2 during the second emergence episode. (a)-(b): N2 move southwestward first. (b)-(c): Next, P2 moves eastward, initiating the collision. The apparent self-separation distance for P2-N2 is shown in Mm with a cyan dashed line. The collisional PIL (yellow) and self-PIL (green) are both presented. The coordinate of the disk center is also given in each panel.}  \label{fig:fig5}
\end{figure}

\begin{figure}
\plotone{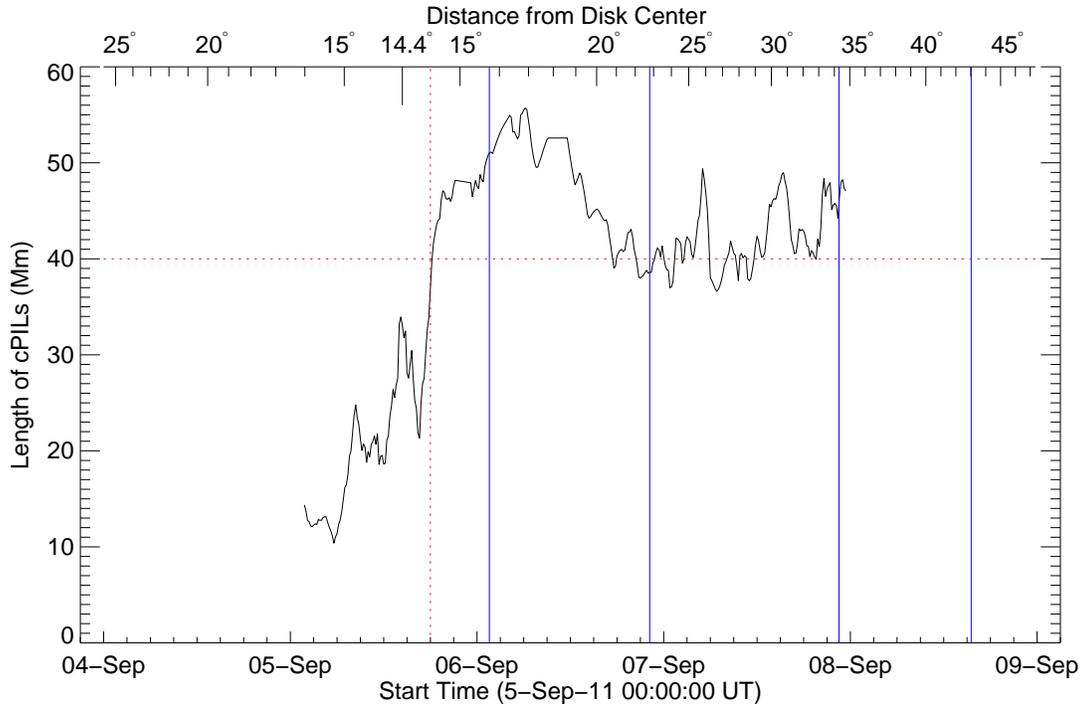}
\caption{Collisional PIL length with time. The red vertical dotted line shows the onset time of collision, which is defined when the continuous length of the collisional PIL definitely exceeds 40 Mm (red horizontal line). The sun-center angle of the emerging region is presented on the top $x$-axis. The onset times of the four eruptions are also presented by blue vertical lines.} \label{fig:fig6}
\end{figure}

\begin{figure}
\centerline{\includegraphics[width=0.6\textwidth,clip=]{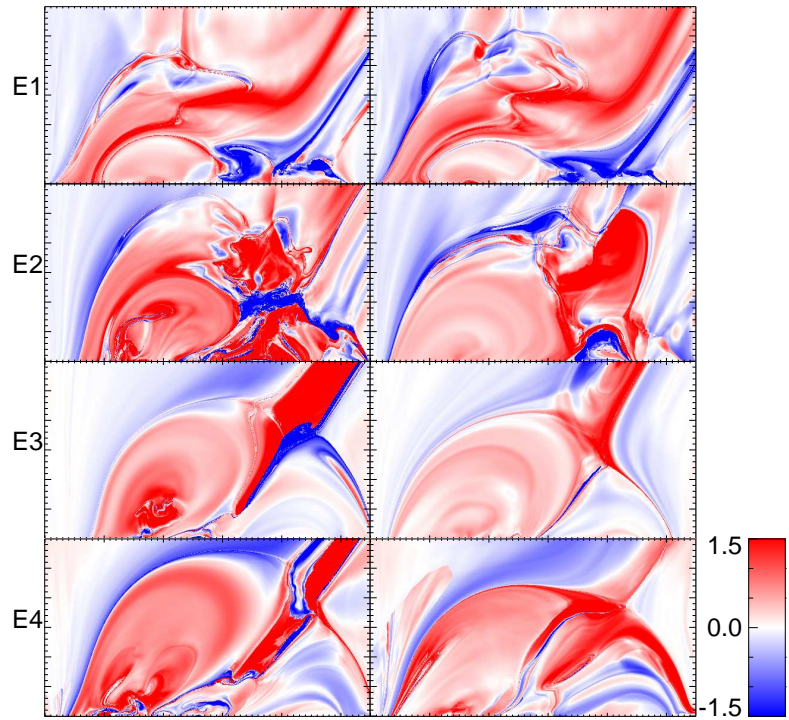}}
\caption{Comparisons of the twist map before (left column) and after the four major eruptions (right column). The cross section is along the white vertical line in Figure 2a. At lower right is the color bar that indicates the twist number.} \label{fig:fig7}
\end{figure}

At the later phase of the second emergence episode, E1 occurrs at 01:35 UT September 6. \citet{2016Jiangcw} indicate that this eruption was triggered by the reconnection between the newly emerging magnetic arcades and the pre-existing ambient fields. The fast flux emergence in their study just corresponds to the second emergence episode. Their numerical model did not produce a fully formed magnetic flux rope, which is consistent with our flux deficit measurement, namely that the collision just began and less nonpotentiality was accumulated along the cPIL. Figure~\ref{fig:fig3} shows that the flux deficit begins to increase after the first eruption and grows quickly after 15:00 UT September 6 when the spot rotation of P2 starts (see Figure~\ref{fig:fig10}c) and both the free energy and volume helicity increase (see Figure~\ref{fig:fig1}a and~\ref{fig:fig1}b). Figure~\ref{fig:fig11}a shows that two bundles of the coronal bright structure observed at the EUV 94 \AA~wavelength at 09:00 UT September 6. During the spot rotation, it seems that the two bundles of bright structures reconnect with each other, and EUV brightenings can be seen in Figure~\ref{fig:fig11}b. Photospheric Doppler velocity field data are also adopted to help investigate magnetic cancellation at a cPIL. The corrected velocity field data are available in the data series $cgem.doppcal\_720s$, which are corrected for the differential rotation and the Solar Dynamics Observatory spacecraft velocity \citep[see Section 4 of][]{2020Fisher}. The convective blueshift is further subtracted from the corrected data, which is saved in the DOPPBIAS keyword of the header file \cite[for a detailed discussion, see][]{2013Welsch}. Figure~\ref{fig:fig11}c shows that the photospheric Doppler velocity field appears in a large patch between P2 and N1$^S$. The increased flux deficit is accompanied by a dominance of average redshifts at the cPIL. The twist maps of Figure~\ref{fig:fig11}d show that two bundles of twisted fields gradually coalesce. A TCR process, and/or the reconnection of an MFR structure with ambient fields or with other weakly twisted MFR structures may happen along the cPIL (due to collisional shearing). They may coalesce into larger and more monolithic structures, as shown by the observations at 94 \AA~wavelength (see the animation of Figure~\ref{fig:fig11}b). The dominant redshift signal at the cPIL is interpreted as being consistent with the submergence of the post-reconnected loops (i.e., flux cancellation). The two bundles of twisted fields should be the MFRs. The formation of the MFRs is also associated with the collisional shearing process. Magnetic flux cancellation occurrs along the cPIL. It canceled the flux of $\sim$0.7 $\times$ 10$^{21}$ Mx with an uncertainty of $\sim$13.2\% from the beginning of September 6 to the third eruption within the 30$^\circ$ sun-center distance. The canceled flux is represented by the flux deficit in Figure~\ref{fig:fig3}. The deficit amounts to 24$\%$ of the total unsigned flux of the bipolar magnetic region before the third major eruption.

\begin{figure}
\plotone{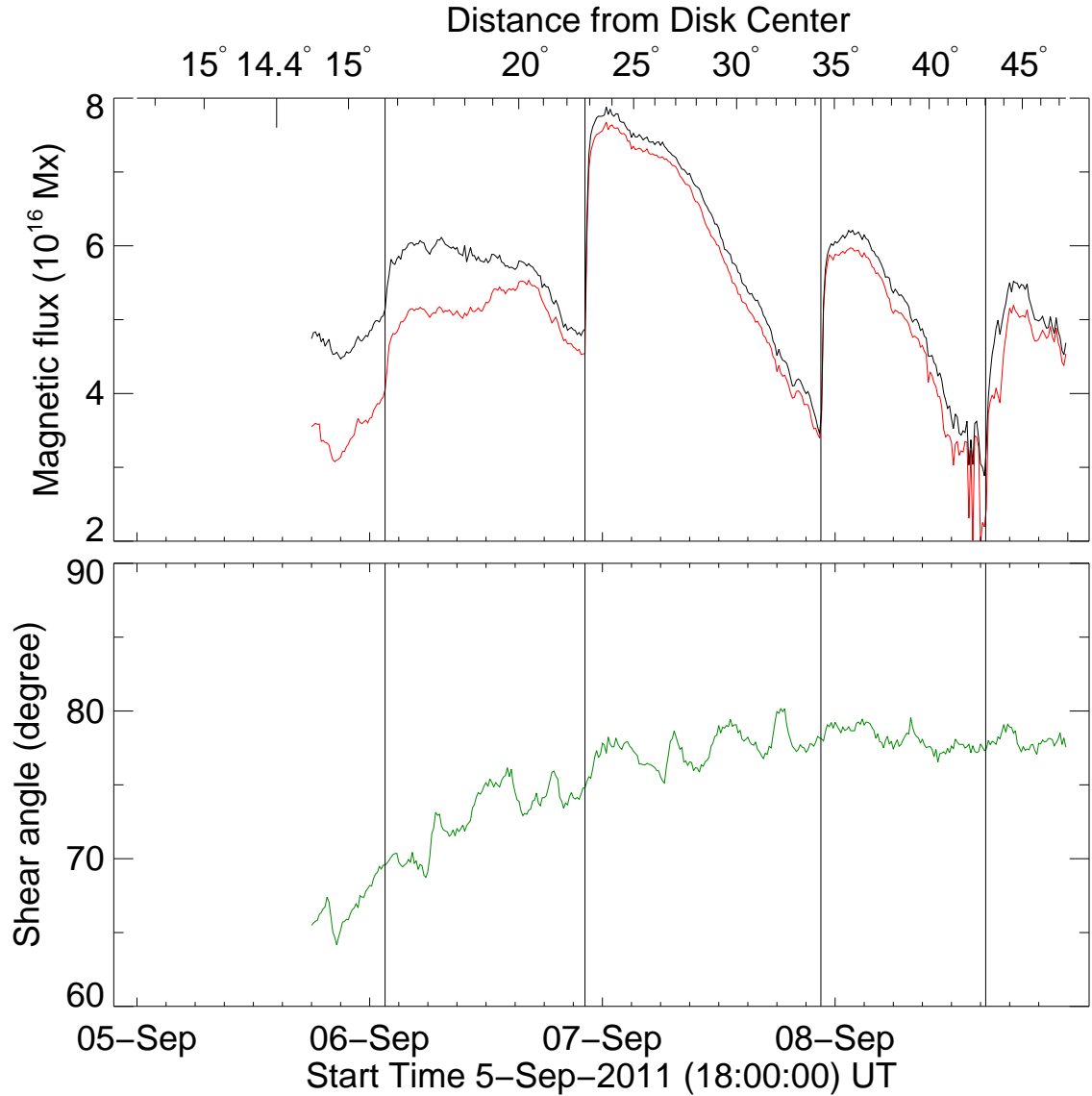}
\caption{Upper panel: integrated tangential magnetic fields B$_t$ (upper black curve) within the green rectangle in Figure 2a. The long side of the rectangle is kept parallel to the PIL. Integrated tangential fields parallel to the PIL (B$_{tpara}$) are plotted in red. Bottom panel: average shear angle within the green rectangle.} \label{fig:fig8}
\end{figure}

\begin{figure}
\plotone{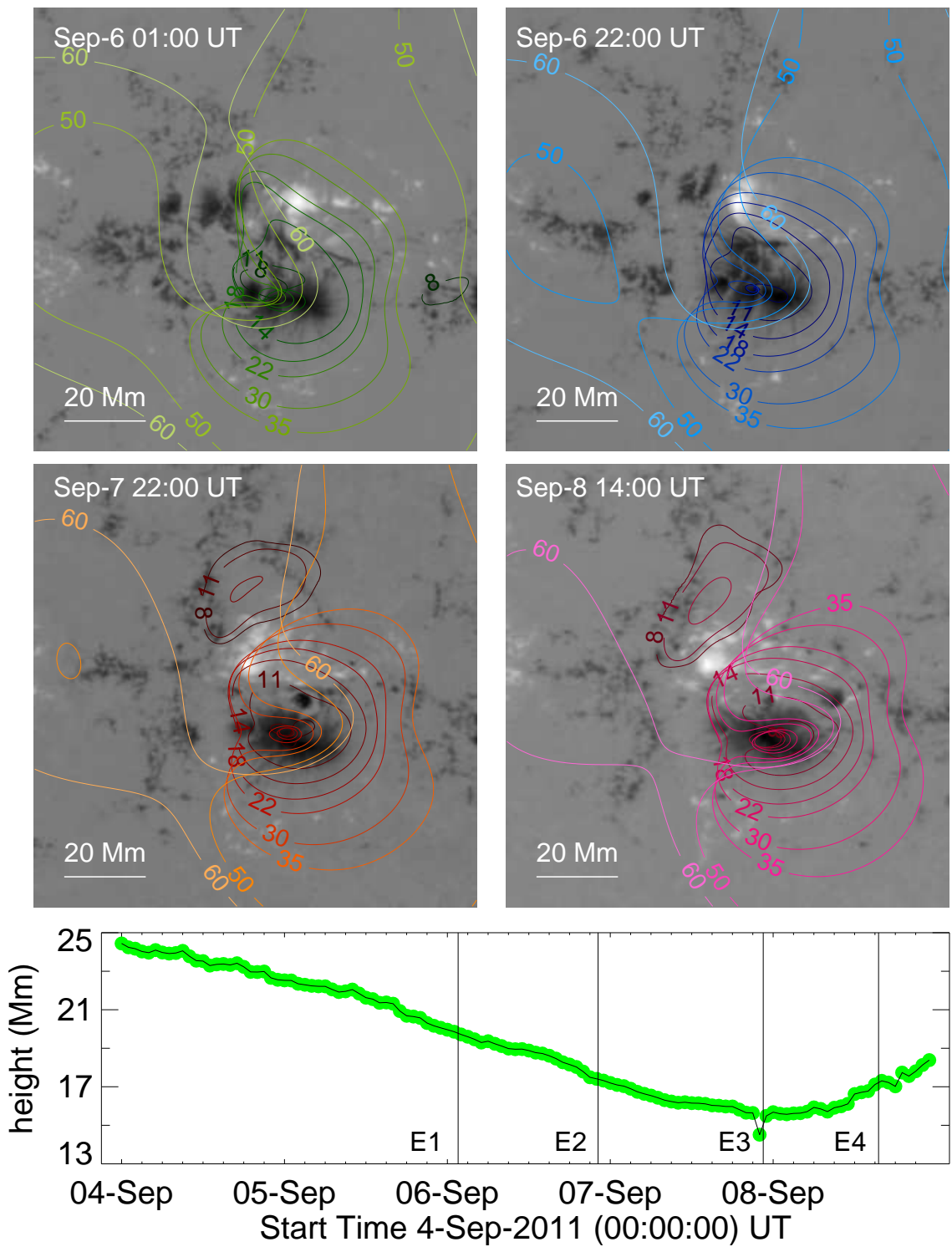}
\caption{Two-dimensional distributions and the time evolution of the decay index above the emerging region. Upper four panels: decay index with height contour plot overlaid on photospheric magnetograms just before each of the eruptions. Color contours correspond to a decay index n = 1.5 at each height (in Mm). The color of the contour lightens as it increases in height. Bottom panel: time evolution of the critical height with average decay index (within the yellow rectangle in Figure 2a) equal to 1.5.} \label{fig:fig9}
\end{figure}

\begin{figure}
\plotone{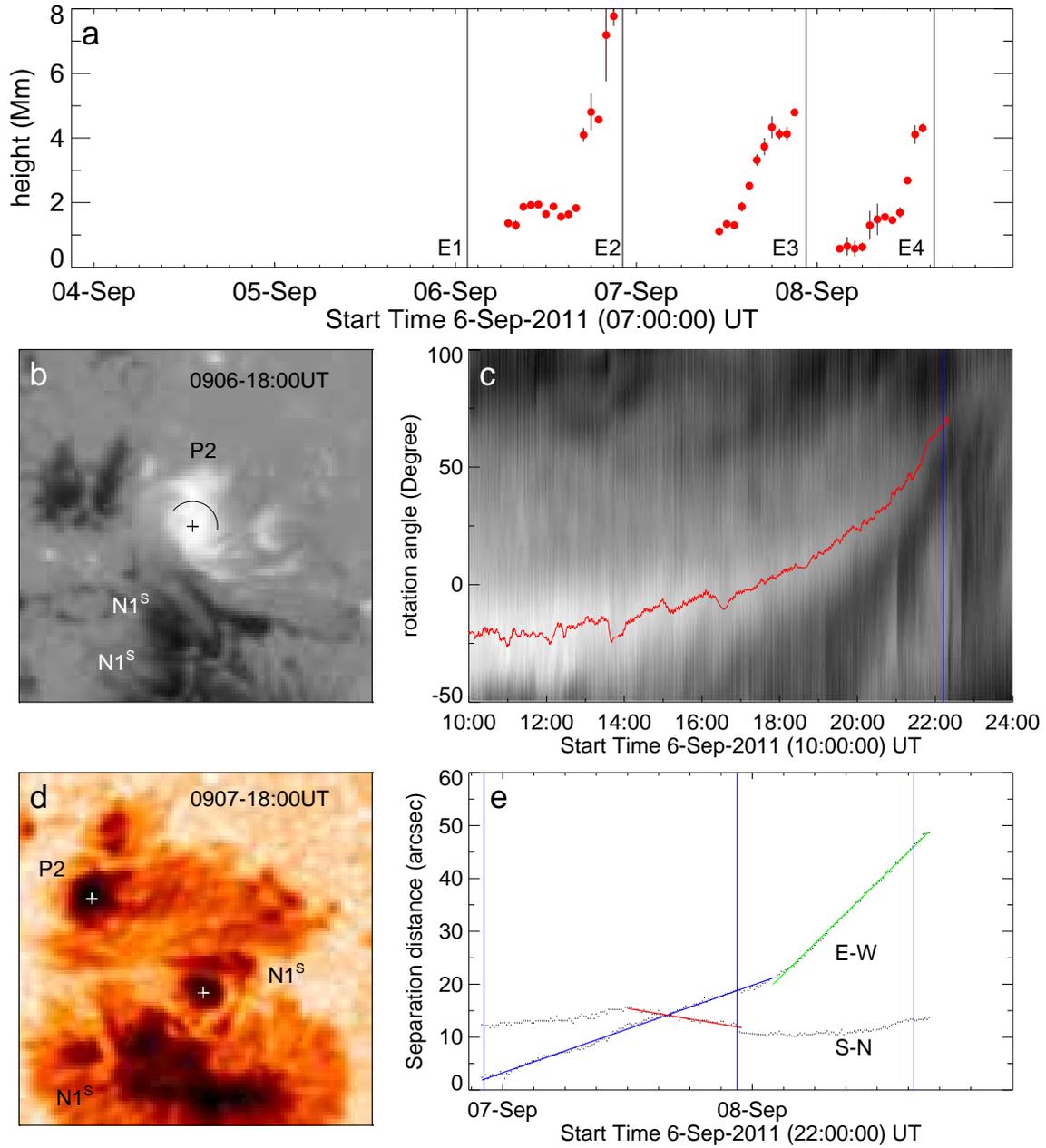}
\caption{Relationship between the height evolution of the MFRs and the photospheric motions. (a) Height evolution of each MFR with uncertainties. (b) HMI magnetogram during the sunspot rotation. An arc slit is centered at the black cross. (c) Stackplot of the slit in (a). The red curve traces the maximum of each slit. The blue vertical line marks the eruption on September 6 (E2). (d) Shearing motions (E-W) and converging motions (S-N) between P2 and N1$^S$ shown in the HMI continuum image, and (e) the related time-distance plot in the two directions. The colored straight lines are the linear fits to the distances (black dotted lines). The vertical lines mark E2, E3, and E4.} \label{fig:fig10}
\end{figure}

\begin{figure}
\plotone{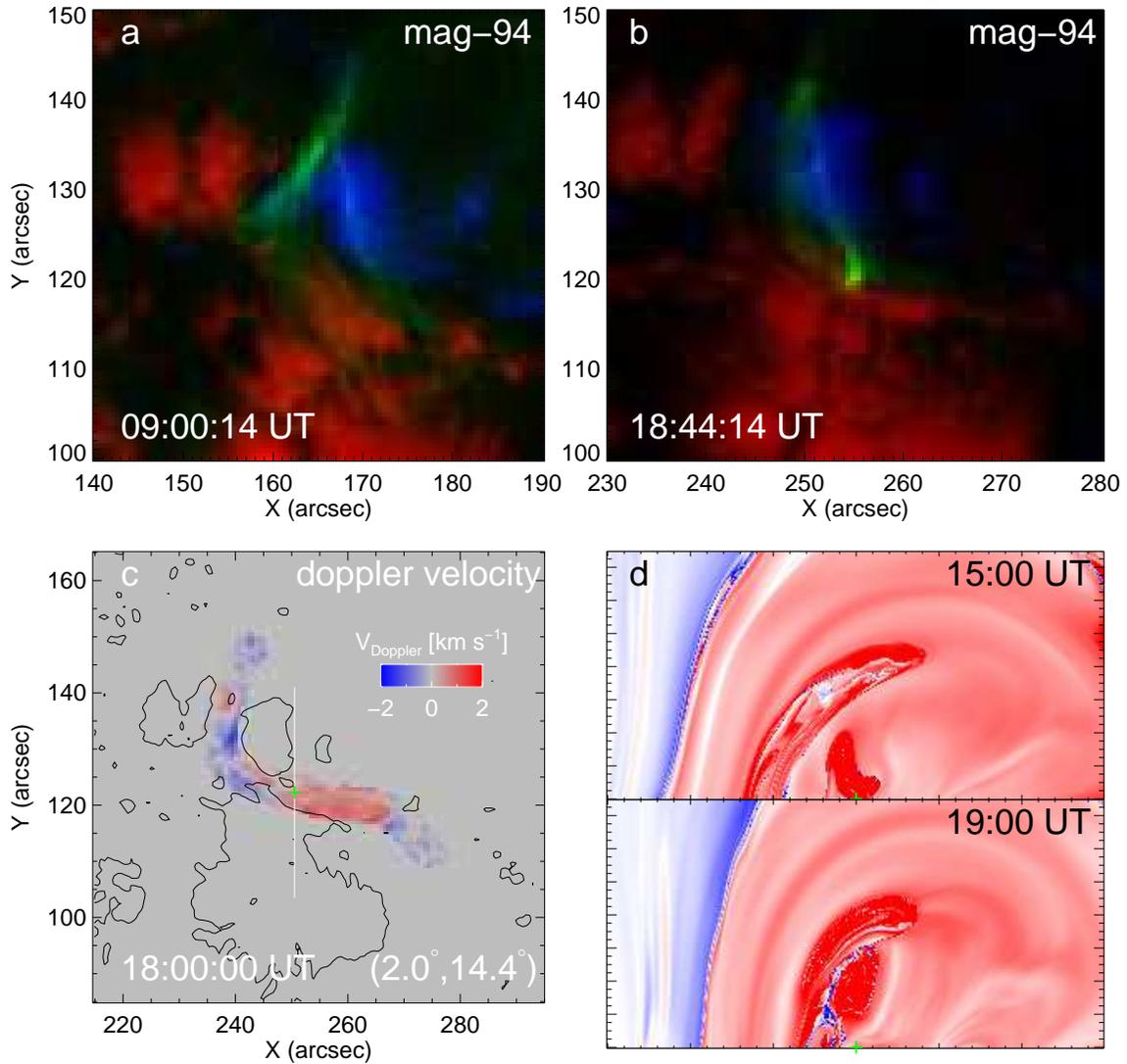}
\caption{(a)--(b): Composite image of 94 \AA~(green), and HMI magnetogram (red/negative, blue/positive) showing plasma emission at 6.0 MK. (c) Doppler velocity measurements at the location of the collisional PIL before the second eruption. The magnetic field contours are overplotted. To improve visual clarity we only show the Doppler velocity along the filament. (d) Twist maps (approximately along the white vertical line in (c)) showing the coalescence of two bundles of magnetic fluxes. The green crosses in (c) and (d) are the reference points for the filament. The color bar is the same as shown in Figure~\ref{fig:fig7}. Animation of (b) is available. The animation starts at 16:59:26 UT and ends at 21:59:38UT. The real-time duration of the animation is 4s.} \label{fig:fig11}
\end{figure}

\begin{figure}
\plotone{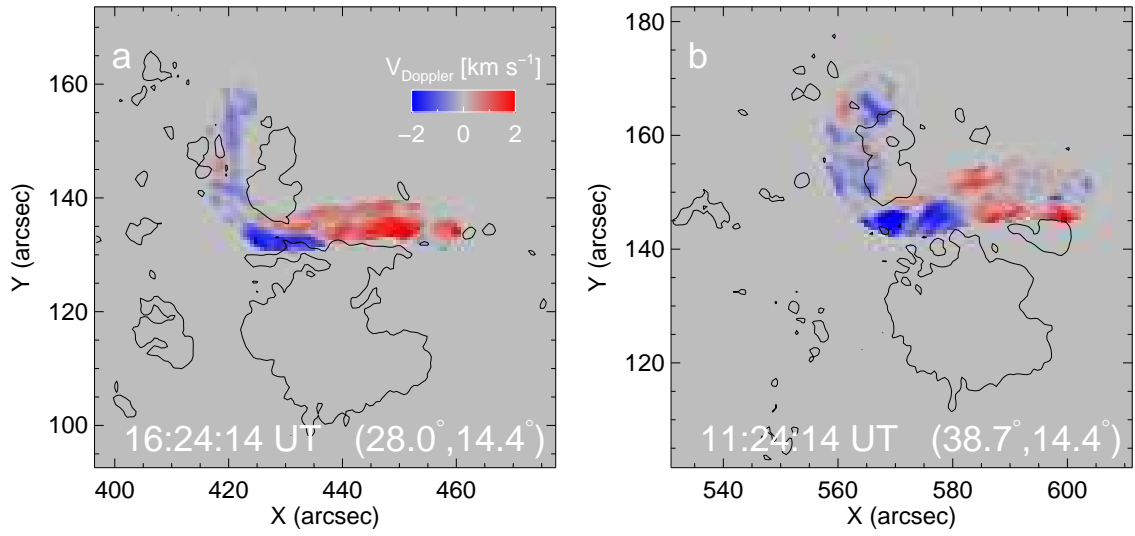}
\caption{Doppler velocity measurements at the location of the collisional PIL before the third (left) and fourth eruptions (right). We also only show the Doppler velocity along the filaments.} \label{fig:fig12}
\end{figure}

It is worth mentioning that the negative parasitic polarity emerged between N1$^N$ and N1$^S$. Its positive counterpart seems continuously to cancel with N1$^N$. It merges with N1$^S$ moving westward at the later phase of September 6. The westward-moving negative polarity N1$^S$ is compressed to an elongated shape by the eastward-moving P2, moving the cPIL into an east-west direction. Figure~\ref{fig:fig8} shows that the average shear angle almost reaches its maximum value on September 7, which implies that the shearing motion becomes strong. Figure~\ref{fig:fig10}e shows that P2 and N1$^S$ separate from each other in the constant speed (blue line), while a converging motion (red line) is ongoing at the later time of September 7. The convergence of the opposite-signed magnetic polarities causes a further cancellation along the cPIL, the amount of which can be estimated through the increase of the flux deficit in Figure 3. Figure~\ref{fig:fig12}a shows that the increase in the flux deficit is also accompanied by the dominance of redshift signals over blueshift signals along the cPIL.

As N1$^S$ and N2 approach to each other when the AR is beyond the angular distance of 30$^\circ$, it becomes difficult to group them on September 8, and the ``grouping'' uncertainty of the flux deficit becomes large. Thus, we do not perform a quantitative estimation of the canceled flux outside of the confidence region. The separation speed between P2 and N1$^S$ is enhanced on September 8 (the green line in Figure~\ref{fig:fig10}e). Meanwhile, Figure~\ref{fig:fig3} shows that the flux emergence gets stronger during this period. The compact cPIL becomes fragmented. The blueshift signal seems to increase in this period. It should be related to the Evershed flow in the weak penumbra (see Figure~\ref{fig:fig12}b); the northeastern part of N1$^S$ along the cPIL became discrete. Nevertheless, the redshift related to the submergence of the magnetic field was still dominates, which implies that the flux is canceling at the cPIL.

We analyze the four major eruptions and the evolution of the related magnetic fields in the AR 11283. We obtain the relatively accurate amount of the canceled flux with an uncertainty of $\sim$13.2\% for the 1$\sigma$ noise level by using the conjugate flux deficit method of \cite{2019Chintzoglou} in a flux emerging region, which further supports the view that successive solar eruptions in multipolar ARs likely occur in a ``collisional shearing'' scenario. The growth of the canceled flux has a good correlation with the energy accumulation before each of the eruptions. Nonpotentiality is repeatedly established through the converging and shearing motions of unconjugated polarities driven by continuous flux emergence. For most of the eruptions, the existence of MFR structures is revealed by the EUV observations and by the NLFFF model. The MFRs are the products of the reconnection between the colliding magnetic fields. The question of whether the collisional shearing scenario is a necessary condition for the generation of homologous eruptions remains and needs further study in future work.

\acknowledgments
We thank the anonymous referee for the constructive comments and suggestions. We also thank Dr. Chintzoglou for helpful information and instructions on the methodology of the calculation of the fluxes and for some other constructive suggestions. The data used here are courtesy of the NASA/SDO HMI and AIA science teams. We thank the HMI science team for making the SHARP vector magnetograms available to the solar community. The research was supported by NSFC under grant 12073032, 41774179, 42004145, the CAS Strategic Priority Program on Space Science (XDA15018500), and the Specialized Research Fund for State Key Laboratories of China.

\clearpage

\clearpage
\clearpage
\clearpage
\clearpage
\clearpage
\clearpage
\clearpage

\clearpage

\clearpage

\clearpage

\clearpage

\clearpage
\bibliography{references}
\bibliographystyle{aasjournal}



\end{document}